\documentclass[conference]{IEEEtran}
\usepackage{amsmath,amsthm}
\usepackage{cite}
\usepackage{graphicx}
\usepackage{epstopdf}
\usepackage{amsfonts,amsmath,amssymb}
\usepackage{cite}
\usepackage{graphicx}
\usepackage{url}
\usepackage{bm}
\usepackage{bbm}
\usepackage{amssymb}

\begin{document}

\title{{CV-MDI  Quantum Key Distribution via Satellite}}
\author{
\IEEEauthorblockN{Nedasadat Hosseinidehaj and Robert Malaney}
\IEEEauthorblockA{School of Electrical Engineering  \& Telecommunications,\\
The University of New South Wales,\\
Sydney, NSW 2052, Australia\\
neda.hosseini@unsw.edu.au, r.malaney@unsw.edu.au}
}

\vspace{-5cm}

\maketitle
\begin{abstract}
In this work we analyze a measurement-device-independent (MDI) protocol to establish continuous-variable (CV) quantum key distribution (QKD) between two ground stations. We assume communication occurs between the  ground stations via satellite over two independent atmospheric-fading channels dominated by turbulence-induced beam wander.
In this MDI protocol the measurement device is the satellite itself, and
the security of the  protocol is analyzed through an equivalent entanglement-based swapping scheme.
 We quantify the positive impact the fading channels can have on the final quantum key rates, demonstrating how the protocol is able to generate a positive key rate even over high-loss atmospheric channels. This is somewhat counter-intuitive given that the same outcome is only possible  in the low-loss regime for a  measurement device centrally positioned in a fiber-optic channel. Our results show that useful space-based quantum key generation rates between two ground stations are possible even when the relay satellite is held by an adversary. The cost in key rate incurred by altering the status of the satellite from trustworthy to untrustworthy is presented.

%Although MDI protocol can be useful when there is no a direct link between the two ground stations,
%in circumstances in which two ground stations are not able to directly communicate with each other it might lead to a lower quantum key generation rate in comparison with some non-MDI protocols, thereby illustrating the trade-off between measurement-device independence and quantum key generation rate.

\end{abstract}

\section{Introduction}
QKD allows two distant parties, Alice and Bob, to generate a secret key (unknown to a potential eavesdropper, Eve) over insecure quantum and classical channels. In  discrete-variable (DV) QKD systems the key information is encoded on the properties of single photons, and detection is realized by  single-photon detectors e.g., \cite{DV1,DV2,DV3}. In CV-QKD  the key information is encoded on the quadrature variables of the optical field, and detection is realized by high-efficiency  homodyne (or heterodyne) detection techniques e.g., \cite{CV1, CV2, thesis, Weedbrook2012, Weedbrook2013}.

QKD schemes are mostly implemented based on point-to-point protocols where
Alice encodes the key information onto  quantum states which are then transmitted over an unsecured quantum channel. At the output of the channel Bob receives and measures the incoming  states to infer the information of the sender. The secure key rate of  point-to-point QKD is highly dependent on the devices utilized at the sender and receiver stations, particularly the measurement devices. Unfortunately, practical detectors are not perfect and their inherent flaws can become the attack focus of an adversary \cite{attack1, attack2}. To alleviate this concern MDI protocols have been proposed \cite{first_MDI1, 1st_DV_MDI_theory}. In MDI protocols both trusted (legitimate) parties transmit quantum states to a third party (assumed untrustworthy) who realizes the measurements. The measurement outcomes will then be used by the legitimate users to establish secure keys independent of the measurement device.  Such procedures are particularly useful in relay systems, where due to channel issues an intermediate relay \emph{must} be used to complete the communication link. MDI protocols have been well analyzed in theory \cite{first_MDI1, 1st_DV_MDI_theory, CV_MDI1, CV_MDI2, CV_MDI3, CV_MDI_sym, Nature}, and experimentally demonstrated with DV systems \cite{fiber_DV1, fiber_DV2, fiber_DV3, fiber_DV4} and CV systems \cite{Nature}.
%CD QKD Existence of a direct link between the two parties is not always possible, however they might be able to communicate through a third party which is not necessarily reliable.

Previous works on MDI protocols have largely focussed on fixed-attenuation channels such as optical fiber \cite{fiber_DV1, fiber_DV2, fiber_DV3, fiber_DV4}. In the deployment of QKD between two remote ground stations a viable option could be the use of an MDI protocol with a low-Earth orbit (LEO) satellite realizing the measurement. However, the turbulent atmosphere between ground and satellite leads to a fading channel, whose nature of course is radically different from the fixed-attenuation channel. Also, the atmospheric fading channel typically incurs very high loss rates. This is somewhat of a concern for the implementation of a MDI protocol since previous works on fixed-attenuation channels clearly demonstrate the non-viability of such a protocol in the high-loss regime.
 Thus, in terms of the quantum key rates, it remains unclear if the atmospheric-fading channels can lead to an effective implementation of a MDI protocol.
%we are interested in knowing if MDI scheme would be useful when there is no a direct link between Alice and Bob located in the two ground stations, but there exists a link between them and a satellite which can be used as an intermediate relay for measurement.

It is the main purpose of this work to provide a quantitative assessment, in terms of the resulting key rates, of a CV-MDI protocol over atmospheric-fading channels.
Among the many contributors to the transmission fluctuation in Earth-to-satellite communications, we will focus here only on the  fluctuations caused by beam wandering - the largest contributor \cite{fso, beamwander, Usenko}. We will analyze the security of our MDI protocol through the use of an equivalent entanglement-based swapping scheme e.g., \cite{swapping1, swapping2}. In doing this we will utilize some of our own previous work on the Gaussian entanglement produced by entanglement swapping over atmospheric channels\cite{Neda1, Neda2, Neda3}. Our key aim is to compare the performance (in terms of the resultant key rates) of our CV-MDI protocol implemented over atmospheric-fading channels  to the corresponding CV-MDI protocol over  fixed-attenuation channels.
We will also explore the cost, in the context of quantum key rates, of moving from the usual mode of assuming a trustworthy satellite to the MDI mode of assuming an untrustworthy satellite.

This paper is organized as follows: In Section II, our CV-MDI protocol is described in detail. In Section III, the simulation results on the performance of the protocol over the atmospheric-fading channels are presented and discussed. Our conclusions are provided in Section IV.

\section{system model and quantum key rate}
 We now describe the implementation of a CV-MDI protocol and outline how to determine the quantum key rates of this protocol for the atmospheric channel.

\subsection{A CV-MDI Protocol}

Let us first introduce some notation on CV states we will need for our discussions of the MDI protocol.
For a single bosonic mode with annihilation and creation operators $\hat a,\,{\hat a^\dag }$, the quadrature operators $\hat q,\hat p$ are defined by
$\hat q = \hat a + \,{\hat a^\dag }\,,\,\,\,\,\,\hat p = i({\hat a^\dag } - \hat a\,)$
which satisfy the commutation relation $\left[ {\hat q,\,\hat p} \right] = 2i$ (here $\hbar=2$).
The vector of quadrature operators for a quantum state with $n$ modes can be defined as
${\hat R_{1, \ldots ,n}} = \left( {{{\hat q}_1},\,{{\hat p}_1}, \ldots ,{{\hat q}_n},{{\hat p}_n}\,} \right)$. Similarly, ${R_{1, \ldots ,n}} = \left( {{{ q}_1},\,{{ p}_1}, \ldots ,{{ q}_n},{{ p}_n}\,} \right)$ is defined for the corresponding quadrature variables.
Gaussian states are completely characterized by the first moment of the quadrature operators $\left\langle {{{\hat R}_{1, \ldots ,n}}} \right\rangle $ and a covariance matrix (CM) $M$, i.e. a matrix of the second moments of the quadrature operators, which can be written as
\begin{eqnarray}\label{CM}
{M_{ij}} = \frac{1}{2}\left\langle {{{\hat R}_i}{{\hat R}_j} + {{\hat R}_j}{{\hat R}_i}} \right\rangle  - \left\langle {{{\hat R}_i}} \right\rangle \left\langle {{{\hat R}_j}} \right\rangle .
\end{eqnarray}
By local unitary operators, the first moment of every two-mode Gaussian state can be set to zero and the CM can be transformed into the following standard form
\begin{eqnarray}\label{general-CM}
{M}= \left( {\begin{array}{*{20}{c}}
A&C\\
{{C^T}}&B
\end{array}} \right),\,
\end{eqnarray}
where
$A = aI\,,\,B = bI\,,\,C = diag\left ( {{c_ + },{c_ - }} \right )$,
$a,b,{c_ + },{c_ - } \in \mathbb{R}$, and
$I$ is a $2 \times 2$ identity matrix.

In a typical point-to-point QKD protocol two distant trusted parties, Alice and Bob, first create two sets of correlated data by exchanging  quantum states over an unsecured quantum channel. In a prepare-and-measure (PM) scheme Alice is the sender who prepares the quantum states and Bob is the receiver who measures the incoming quantum states. Following the steps of reconciliation and privacy amplification over a public (but authenticated) classical channel, Alice and Bob can subsequently generate a secret key even in the presence of Eve.
%However, there is a problem with conventional point-to-point QKD systems.

In this work we will largely focus on the CV-MDI protocol \cite{CV_MDI1, CV_MDI2, CV_MDI3, CV_MDI_sym, Nature} whose sources are Gaussian states,
% coherent or squeezed states, randomly modulated by a Gaussian distribution.
experimentally demonstrated over free-space (non-fading) channels \cite{Nature}.

\subsubsection{CV-MDI protocol in the PM Scheme}
The CV-MDI protocol in the PM scheme proceeds as follows:\break
\emph{Preparation:} If squeezed states represent the initial quantum resource, Alice (Bob) prepares mode~$A$ $(\hat q_A,\hat p_A)$ (mode~$B$ $(\hat q_B,\hat p_B)$) in a squeezed state with CM ${M_s} = diag(1/v,v)$, where $v = \exp (2{r_s})$, and where $r_s$ is the single-mode squeezing. The squeezed quadrature of mode~$A$ (mode~$B$) is then modulated  by a random Gaussian-distributed variable with zero mean and variance $v_m$ such that $v_m=v-1/v$. We will assume modes~$A$ and $B$ are modulated by the same but independent Gaussian distributions. Choosing to squeeze either the $\hat q$ or $\hat p$ quadrature is based on a random bit generated at Alice and Bob's side.

If coherent states represent the initial quantum resource, Alice (Bob) prepares mode~$A$ $(\hat q_A,\hat p_A)$ (mode~$B$ $(\hat q_B,\hat p_B)$) in a coherent state, where each quadrature of mode~$A$ and mode~$B$ are independently modulated by a random Gaussian-distributed variable with zero mean and variance $v'_m$.
%We will again assume modes~$A$ and $B$ are modulated by the same but independent Gaussian distributions of variance $v_m$.
  \break
\emph{Transmission:} Alice and Bob transmit modes~$A$ and $B$ over the insecure lossy channels to the untrusted relay. \break
\emph{Measurement:} A CV Bell measurement is performed on the incoming modes $A''$ and $B''$ (where the $''$ indicates that the states have now incurred losses), which means modes $A''$ and $B''$ are combined on a balanced beam splitter whose output ports are conjugately homodyned. As a result, the quadrature operators ${\hat q_-} = \frac{1}{{\sqrt 2 }}\left( {{{\hat q}_{A''}} - {{\hat q}_{B''}}} \right)$ and ${\hat p_+} = \frac{1}{{\sqrt 2 }}\left( {{{\hat p}_{A''}} + {{\hat p}_{B''}}} \right)$ are measured by the two homodyne detectors, and the classical measurement outcomes $(q_-,p_+)$  with probability $p\left( {{{q_-}},{{p_+}}} \right)$ are then communicated over a public channel to Alice and Bob. Then Bob modifies his data based on the Bell measurement outcomes, while Alice keeps her data unchanged. As a result, the mutual information between Alice and Bob becomes nonzero, and a correlation is created between the two parties. As the relay may be controlled by Eve, she does know the measurement results, however, this knowledge would not help her extract precise information on Alice and Bob's encodings. \break
\emph{Post-processing:} After the establishment of a sufficiently large amount of correlated data, Alice and Bob proceed with the classical post-processing over an authenticated public channel which starts by applying sifting process (no sifting is needed when the two partners prepare the coherent states) and continues with parameter estimation, information reconciliation and privacy amplification to distill a secret key. The reconciliation can be performed in two ways; either Alice's data or Bob's data are the reference. Any false reporting of the Bell measurement results by the relay will be readily detected in the post-processing phase.

Although the CV-MDI protocol is practically implemented in a PM scheme, we will study the protocol in an equivalent entanglement-based (EB) scheme that invokes CV-entanglement swapping at the satellite.
\subsubsection{CV-MDI protocol in the EB Scheme}
In the EB scheme, a pair of two-mode squeezed vacuum (TMSV) states $\rho_{12}$ and $\rho_{34}$ with the same two-mode squeezing $r_t \in \left[ {0,\,\infty } \right)$ are initially owned by Alice and Bob respectively, where indices~$1-4$ indicate the  modes. Let us consider the initial two-mode Gaussian entangled states $\rho_{12}$ and $\rho_{34}$ having zero mean and CM of the following form
\begin{eqnarray}\label{initial}
M_i= \left( {\begin{array}{*{20}{c}}
{v\,I}&{\sqrt {{v^2} - 1} \,Z}\\
{\sqrt {{v^2} - 1} \,Z}&{v\,I}
\end{array}} \right) ,
\end{eqnarray}
where $Z = diag\left( {1, - 1} \right)$, and $v = \cosh \left( {2r_t} \right)$ is the quadrature variance of each mode. Modes~1 and 4 are held by Alice and Bob, while modes~2 and 3 are transmitted towards the intermediate relay over the insecure lossy channels with transmissivities of $\tau_A$ and $\tau_B$, respectively.
%Transmission of mode~2 (mode~3) over the lossy channel can be simulated by the interaction of mode~2 (mode~3) and a vacuum noise in a beam splitter with the transmission efficiency $\tau_A$ ($\tau_B$) in Alice's link (Bob's link).
Assuming the fixed values of $\tau_A$ and $\tau_B$, the received states at the relay, $\rho_{1{2''}}$ and $\rho_{{3''}4}$, are still Gaussian, and described with CMs
\begin{equation}\
\begin{array}{l}
{M_{1{2''}}} = \left( {\begin{array}{*{20}{c}}
{v\,I}&{\sqrt {{\tau _A}} \sqrt {{v^2} - 1} \,Z}\\
{\sqrt {{\tau _A}} \sqrt {{v^2} - 1} \,Z}&{\left( {{\tau _A}v + (1 - {\tau _A}) + {\varepsilon _A}} \right)\,I}
\end{array}} \right)\\
\\
{M_{{3''}4}} = \left( {\begin{array}{*{20}{c}}
{\left( {{\tau _B}v + (1 - {\tau _B}) + {\varepsilon _B}} \right)\,I}&{\sqrt {{\tau _B}} \sqrt {{v^2} - 1} \,Z}\\
{\sqrt {{\tau _B}} \sqrt {{v^2} - 1} \,Z}&{v\,I}
\end{array}} \right),
\end{array}
\label{transmitted}
\end{equation}
where ${{\varepsilon _A}}$ and ${{\varepsilon _B}}$ are the excess noise contributions assumed to be independent of the channel loss.
In implementations of CV-MDI QKD,  excess noise generally comes from the two sources; quantum state preparation at the transmitters (laser's phase noise and imperfections in the modulation), and Bell measurement at the receiver (electronic noise and imperfections in the homodyne detectors). We will assume the transmitter noise is negligible relative to the receiver noise.

The received modes~$2''$ and $3''$ are swapped via a Bell measurement at the intermediate relay, where the modes~$2''$ and $3''$ are combined in a balanced beam-splitter, yielding output modes~$s$ and $t$. Then, the new quadratures ${\hat q_s} = \frac{1}{{\sqrt 2 }}\left( {{{\hat q}_{2''}} - {{\hat q}_{3''}}} \right)$ and ${\hat p_t} = \frac{1}{{\sqrt 2 }}\left( {{{\hat p}_{2''}} + {{\hat p}_{3''}}} \right)$ are measured by two homodyne detectors, providing the classical outcomes $\left( {{{q_s}},{{p_t}}} \right)$ with probability $p\left( {{{q_s}},{{p_t}}} \right)$.
The swapping process continues with the relay communicating the Bell measurement results through a classical public channel to Alice and Bob. After the swapping process the CM of the conditional Gaussian state $\rho_{14}$ shared between Alice and Bob can be obtained \cite{Obermaier, Neda1}
\begin{eqnarray}\label{conditional}
\begin{array}{l}
{M_{14}} = \left( {\begin{array}{*{20}{c}}
{vI}&0\\
0&{vI}
\end{array}} \right) - \,\\
\\
\,\,\,\,\,\,\,\,\,\,\,\,\,\,\,\,\,\left( {{v^2} - 1} \right)\left( {\begin{array}{*{20}{c}}
{\frac{{{\tau _A}}}{\theta }I}&{ - \frac{{\sqrt {{\tau _A}{\tau _B}} }}{\theta }Z}\\
{ - \frac{{\sqrt {{\tau _A}{\tau _B}} }}{\theta }Z}&{\frac{{{\tau _B}}}{\theta }I}
\end{array}} \right),
\end{array}
\end{eqnarray}
where $\theta  = (v - 1)({\tau _A} + {\tau _B}) + ({\varepsilon _A} + {\varepsilon _B}) + 2$. After receiving the Bell measurement outcomes, Bob displaces mode~4 and obtains mode~$4'$, while Alice keeps mode~1 unchanged. Then Alice and Bob measure modes~1 and $4'$ by homodyne (or heterodyne) detectors to create correlated data.
After the establishment of a sufficiently large amount of correlated data, Alice and Bob proceed with the classical postprocessing over an authenticated public channel to create a secret key.

\subsubsection{Equivalence of CV-MDI protocol in the EB Scheme and the PM Scheme}
(See Appendix for more details.) In the EB scheme of the CV-MDI protocol, if Alice (Bob) applies a homodyne detection on mode $1$ (mode $4$) of the initial TMSV state $\rho_{12}$ ($\rho_{34}$), the prepared state in its equivalent PM scheme is a squeezed state on mode $2$ (mode $3$), whose squeezed quadrature is modulated by a random Gaussian distributed variable of zero mean and variance $v_m=v-1/v$. If Alice (Bob) applies a heterodyne detection on mode~1 (mode 4), the prepared state in its equivalent PM scheme is a coherent state on mode~2 (mode 3), whose both quadratures are independently modulated by a random Gaussian-distributed variable of zero mean and variance $v'_m=v-1$.

Note that these homodyne and heterodyne detections by Alice and Bob in the EB scheme can be postponed to after the Bell measurement at the relay since the local measurements commute. In the protocol of coherent states, the displacement variance $v'_m$ is the result of a feasible modulation, hence $v'_m$ can reach high values, for instance $v'_m=60$ \cite{Nature}. On the contrary, in the protocol of squeezed states, high values of squeezing are experimentally challenging to achieve, hence the value of $v$ and also $v_m$ is practically limited. For example, $v=5.05$ is equivalent to the two-mode squeezing of 10dB (the state-of-the-art vacuum squeezing for the TMSV state).

\subsection{Quantum Key Rate}

Having the conditional Gaussian state with the CM in the form of Eq.~\eqref{conditional}, we are able to analytically compute the key rate $K$ of the protocol in bits per pulse,
%\footnote{Here, we assume the pulse generation rate is the same at Alice and Bob's side, and also each pulse sent from Alice and Bob is measured at the relay. Note the key rate in units of bits per second can be calculated as multiplying the key rate in bits per pulse by the pulse generation rate.}
under the assumption of Gaussian attacks by Eve \cite{G-attck}.
%\footnote{The key rate calculated in this work is a lower bound on the key rate since we assume a Gaussian attack (the most effective attack) in our security analysis.}
We will make the usual assumption that the number of exchanges between the trusted parties and the relay are considered infinite.
%Gaussian transmitted states from both trusted parties towards the intermediate relay are in the same type; coherent states or squeezed states.
% In the entanglement-based representation, transmission of coherent states (squeezed states) towards the intermediate relay is equivalent to apply the heterodyne detection (homodyne detection) on modes 1 and 4, however these detections can be postponed after Bell measurement since local measurements commute.
 We will also assume Alice and Bob before their detections share a Gaussian state $\rho_{AB}$ having zero mean value and CM in the following form
\begin{equation}\label{general}
M_{AB} = \left( {\begin{array}{*{20}{c}}
{a\,I}&{c\,Z}\\
{c\,Z}&{b\,I}
\end{array}} \right).
\end{equation}
Note that the CM of the conditional Gaussian state of the MDI protocol, i.e., $M_{14}$ is also in the above form. Assuming a perfect reconciliation algorithm, the key rate is given by $K = I_{AB} - I _{E}$\footnote{In a realistic reconciliation algorithm, Alice and Bob obtain only a fraction of the mutual information $I_{AB}$, and thus the key rate is given by $K = \xi {I_{AB}} - {I_E}$ where $\xi $ is the reconciliation efficiency.}, where $I_{AB}$ is the mutual information between Alice and Bob \cite{thesis, Weedbrook2012, Weedbrook2013}. In the reconciliation step, If Alice's data are the reference, $I_{E} = I _{AE}$ is Eve's mutual information with Alice, while if Bob's data are the reference, $I_{E} = I _{BE}$ is Eve's mutual information with Bob. Note that $I_{AB}$ is the same regardless of whose data are the reference of reconciliation.

 Recalling that we are considering two protocols of the EB scheme in which the trusted parties apply the same type of detection  (homodyne detection or heterodyne detection to their own modes), we can now proceed to determine key rates for each case.

\emph{(i) Homodyne detection by the trusted parties:} Alice and Bob's mutual information is given by ${I_{AB}} = \frac{1}{2}{\log _2}\left( {\frac{a}{{a - \left( {{c^2}/b} \right)}}} \right)$. If Alice is the reference in the reconciliation step, Eve's mutual information with Alice can be calculated as ${I_{AE}} = S({\rho _E}) - S({\rho _{E\left| A \right.}})$, where $S({\rho _E})$  is the von Neumann entropy of Eve's state before Alice and Bob's detections,  and $S({\rho _{E\left| A \right.}})$  is the von Neumann entropy of Eve's state conditioned on Alice's detection. Note that because Eve provides a purification of Alice and Bob's density matrix, we can write $S({\rho _E}) = S({\rho _{AB}})$, and $S({\rho _{AB}})$ can be calculated through the symplectic eigenvalues ${\nu _\pm}$ of $M_{AB} $ as $S({\rho _{AB}}) = f({\nu _+}) + f({\nu _-})$, where $f(x) = \frac{{x + 1}}{2}{\log _2}\left( {\frac{{x + 1}}{2}} \right) - \frac{{x - 1}}{2}{\log _2}\left( {\frac{{x - 1}}{2}} \right)$, and  $\nu _ \pm ^2 = \left( {\Delta  \pm \sqrt {{\Delta ^2} - 4\det \left( {{M_{AB}}} \right)} } \right)/2$, with $\Delta  = {a^2} + {b^2} - 2{c^2}$. Next, $S({\rho _{E\left| A \right.}}) = f(\nu )$ where ${\nu ^2} = b\left( {b - \left( {{c^2}/a} \right)} \right)$.

If Bob is the reference in the reconciliation step, we are required to calculate Eve's mutual information with Bob which is given as ${I_{BE}} = S({\rho _E}) - S({\rho _{E\left| B \right.}})$, where $S({\rho _E})$ is calculated in the same way as before, while $S({\rho _{E\left| B \right.}}) = f(\nu )$ where ${\nu ^2} = a\left( {a - \left( {{c^2}/b} \right)} \right)$.

\emph{(ii) Heterodyne detection by the trusted parties:} Alice and Bob's mutual information is given by ${I_{AB}} = {\log _2}\left( {\frac{{b + 1}}{{b + 1 - \left( {{c^2}/(a + 1)} \right)}}} \right)$. If Alice is the reference in the reconciliation step, Eve's mutual information with Alice ${I_{AE}}$ is calculated in the same way as the homodyne detection protocol (where Alice is the reference) except here $\nu  = b - \left( {{c^2}/(a + 1)} \right)$.

If Bob is the reference in the reconciliation step, Eve's mutual information with Bob ${I_{BE}} $ is calculated in the same way as the homodyne detection protocol (where Bob is the reference) except here $\nu  = a - \left( {{c^2}/(b + 1)} \right)$.

%Note that in the symmetric setting i.e., $\tau_A=\tau_B$,
%% where the relay is exactly located in the middle,
% the key rate will be the same no matter either Alice or Bob is the reference for the reconciliation.

%providing both parties apply the same type of detection on their own modes.
\subsection{CV-MDI Protocol over Atmospheric-Fading Channels}
%Previous works on the MDI protocols have mostly focussed on the fixed attenuation channels, such as optical fiber. Recently, the CV-MDI protocol has been experimentally demonstrated over the free-space links \cite{Nature}. However, the current fiber and the free-space links are not able to implement a real global-scale QKD system. Extension of the fiber links beyond few hundred kilometers is challenging, and the terrestrial free-space links are extremely sensitive to the external environment such as the objects interposed in the line of sight.

In this work we study a satellite-based communication scheme to implement the CV-MDI protocol between two remote ground stations. Let us assume Alice and Bob are located in the ground stations, and there exists a direct communication link from each ground station to the satellite, with the satellite acting as an intermediate relay (untrusted relay) that applies the Bell measurement to the incoming quantum states.

In atmospheric channels fluctuations in the transmissivity $\eta_{tran}$ can be the result of several effects. Such fading channels can be characterized by a distribution of values $\eta$ with a probability density distribution $p\left( \eta  \right)$, where $\eta=\sqrt{\eta_{tran}}$. As  in other recent studies \cite{fso, beamwander, Usenko}, we will assume that atmospheric fading is solely due to beam wander. Assuming the beam spatially fluctuates around the centroid of the receiver's aperture, the probability density distribution $p\left( \eta  \right)$ can be described by the log-negative Weibull distribution \cite{beamwander},
%$$
\begin{equation}\
p\left( \eta  \right) = \frac{{2{L^2}}}{{\sigma _b^2\gamma_s \eta }}{\left( {2\ln \frac{{{\eta _0}}}{\eta }} \right)^{\left( {\frac{2}{\gamma_s }}- 1 \right) }}\exp \left( { - \frac{{{L^2}}}{{2\sigma _b^2}}{{\left( {2\ln \frac{{{\eta _0}}}{\eta }} \right)}^{\left( {\frac{2}{\gamma_s }} \right)}}} \right)
\label{f1}
\end{equation}
%$$
for $\eta  \in \left[ {0,\,{\eta _0}} \right]$, with $p\left( \eta  \right) = 0$ otherwise.
Here, $\sigma _b^2$ is the beam wander variance,
 $\gamma_s$ is the shape parameter,  $L$ is the scale parameter, and ${\eta _0}$ is the  maximum value of $\eta$. The latter three parameters are given by
%$$
\begin{eqnarray}\label{f2}
\begin{array}{l}
\gamma_s  = 8h\frac{{\exp \left( { - 4h} \right){I_1}\left[ {4h} \right]}}{{1 - \exp \left( { - 4h} \right){I_0}\left[ {4h} \right]}}{\left[ {\ln \left( {\frac{{2\eta _0^2}}{{1 - \exp \left( { - 4h} \right){I_0}\left[ {4h} \right]}}} \right)} \right]^{ - 1}}\\
\\
L = \beta{\left[ {\ln \left( {\frac{{2\eta _0^2}}{{1 - \exp \left( { - 4h} \right){I_0}\left[ {4h} \right]}}} \right)} \right]^{ - \left( {{1 \mathord{\left/
 {\vphantom {1 \gamma_s }} \right.
 \kern-\nulldelimiterspace} \gamma_s }} \right)}}\\
\\
\eta _0^2 = 1 - \exp \left( { - 2h} \right) ,
\end{array}
\end{eqnarray}
%$$
where ${I_0}\left[ . \right]$ and ${I_1}\left[ . \right]$ are the modified Bessel functions, and where $h = {\left( {{\beta \mathord{\left/
 {\vphantom {a W}} \right.
 \kern-\nulldelimiterspace} W}} \right)^2}$, with $\beta$ being the receiver aperture radius and $W$ the beam-spot radius. In our  calculations we will adopt $W=\beta$, and let  adjustments to the value of $\sigma _b$ set the mean fading loss.
 We can ignore the effects of dephasing in the atmospheric channel\cite{sem}. We will  assume
 a local oscillator passing through the channel (in an orthogonal polarized mode to the signal) allows us to measure the channel transmission factor in real time.

In the CV-MDI protocol over the atmospheric channels, one mode of each initial entangled state is kept by the ground station and the second mode of each state is transmitted to the
satellite through a fading uplink, where the fading uplink from Alice (Bob) to satellite is characterized by transmission coefficient $\eta_A$ ($\eta_B$) and probability density distribution ${p_A}\left( {{\eta _A}} \right)$ (${p_B}\left( {{\eta _B}} \right)$). We will assume that the two fading channels are independent and not necessarily identical. Let the difference between the two fading channels be controlled only by the difference in the value of $\sigma _b$, such that  ${\sigma _{b\_A}} = k{\sigma _{b\_B}}\,,\,k > 0$, where $\sigma _{b\_A}$ ($\sigma _{b\_B}$) is the beam wander standard deviation of Alice's fading uplink (Bob's fading uplink). Note that $k$ allows us to parameterize the difference between the two fading uplinks in terms of the geometry, such as the distance from each ground station to the satellite. After applying the Bell measurement at the satellite and broadcasting the measurement results to the ground stations, the conditional state $\rho_{14}$ after each realization of $\eta _A$ and $\eta _B$ is still Gaussian, and can be completely described by the CM in Eq.~\eqref{conditional}, where the transmissivities $\tau_A$ and $\tau_B$ need to be replaced by $\eta _A^2$ and $\eta _B^2$. Thus, after each realization of $\eta _A$ and $\eta _B$ the key rate can be calculated through the use of the CM of the conditional Gaussian state as $K({\eta _A},{\eta _B})$. Then, the final key rate can be given by averaging over all the possibilities of the two fading channels as $K = \int_0^{{\eta _0}} {\int_0^{{\eta _0}} {K({\eta _A},{\eta _B})p({\eta _A})p({\eta _B})\,d} } {\eta _A}\,d{\eta _B}$.

\section{Simulation Results}

We now simulate the performance (in terms of the quantum key rates) of the CV-MDI protocol over the fading channels.

In Fig.~1(a), we first plot the key rate $K$ (in bits per pulse) for the symmetric setting of the MDI protocol, i.e., ${\sigma _{b\_A}} = {\sigma _{b\_B}}=\sigma _{b}$ as a function of channel loss
 %is plotted for fading channels (top) and fixed attenuation channels (bottom)
for the two types of detection, homodyne detection and heterodyne detection.
This figure shows the key rate when the quadrature variance of the TMSV states is $v=60$ (the same for homodyne and heterodyne detections),  the excess noise terms are the same as ${\varepsilon _A} = {\varepsilon _B} = 0.02$, and Alice's data are the reference of reconciliation.
The abscissa corresponds to $ - 10{\log _{10}}\left( {\int_0^{{\eta _0}} {\eta _A^2{p_A}({\eta _A})d{\eta _A}} } \right)\left( {\int_0^{{\eta _0}} {\eta _B^2{p_B}({\eta _B})d{\eta _B}} } \right)$ and represents the total mean fading losses in the two fading channels.
%(or the total losses in the two fixed attenuation channels)
%under different channels (different $\sigma _{b}$).

 From Fig.~1(a) it is evident the protocol of homodyne detection is able to generate much higher key rates than the protocol of heterodyne detection. Even if the quadrature variance of the TMSV state, $v$, in the protocol of homodyne detection is chosen much lower, for example, when $v=5$ the key rate resulting from the homodyne detection is still significantly higher than the heterodyne detection\footnote{However, in a practical realization of the CV-MDI protocol where the PM scheme is mostly utilized, the generation of squeezed states (equivalent to the homodyne detection protocol) is  more difficult relative to the generation of coherent states (equivalent to the heterodyne detection protocol).}.
Fig.~1(a) has only been plotted for the case where Alice's data are the reference of reconciliation; however, in the symmetric setting, where the two fading channels are characterized with the same probability density distribution, the key rate is always the same no matter whose data is the reference of reconciliation.

We also simulate the performance of the CV-MDI protocol over the fixed-attenuation channels in relation to the fading channels.
In order to make a valid comparison, we  assume the loss in each fixed-attenuation channel is the same as the mean fading loss in the corresponding fading channel, i.e.,  ${\tau _A} = \int_0^{{\eta _0}} {\eta _A^2{p_A}({\eta _A})\,d} {\eta _A}$ and ${\tau _B} = \int_0^{{\eta _0}} {\eta _B^2{p_B}({\eta _B})\,d} {\eta _B}$. Fig.~1(b) shows the key rate (in bits per pulse) over the fixed-attenuation channels as a function of total channel losses, i.e., $ - 10{\log _{10}}({\tau _A}{\tau _B})$ with all the settings and parameters being the same as the corresponding protocols in Fig.~1(a).

The main point we wish to draw from these results is while the CV-MDI protocol is only feasible over  low-loss fixed-attenuation channels, the same protocol is able to achieve positive key rates  over  high-loss fading channels.
 This advantage originates from the stochastic nature of a fading channel.

% In fact, in the high-loss fading channels, although the probability of the large transmission factors is remarkably reduced, it still remains non-zero which leads to the generation of the positive quantum key rates.

%Note in the CV-MDI protocol over the fixed attenuation channels we are also able to increase the quantum key rates using the asymmetric setting in which the party having lower channel loss is chosen as the reference of reconciliation (here Alice). For instance, by setting $\tau_A=0.98$ \cite{Nature} in the heterodyne detection protocol, we are able to achieve the positive key rates for losses up to around 5dB.

%Considering the assumption ${\sigma _{b\_A}} = 0.1{\sigma _{b\_B}}$ in the asymmetric setting, the Alice's link is better, in terms of loss (having less loss), than the Bob's link
%which leads to the generation of higher key rate in the case Alice is the reference of reconciliation. Thus, the optimal setup for MDI protocol in terms of key rate is the asymmetric setup, in which the trusted party having the better link (lower loss) is chosen as a reference for reconciliation.

 \begin{figure}[!t]
    \begin{center}
   {\includegraphics[width=3.4 in, height=3.1 in]{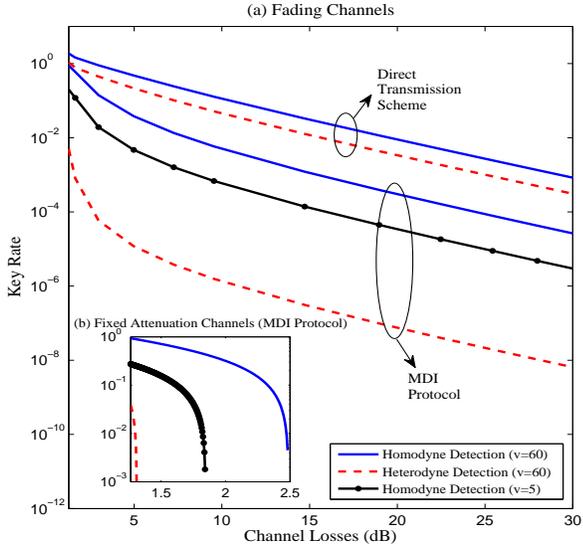}}
    \caption{The key rate (in bits per pulse) resulting from the MDI protocol and from the direct transmission scheme over (a) the fading channels and (b) the fixed attenuation channels using the homodyne detections (solid lines) and the heterodyne detections (dashed lines) in the symmetric setting.}\label{fig:1}
    \end{center}
\end{figure}

\begin{figure}[!t]
    \begin{center}
   {\includegraphics[width=3.4 in, height=3.1 in]{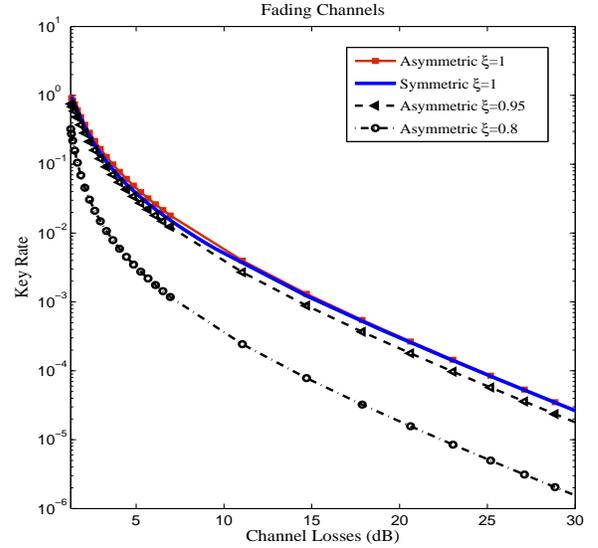}}
    \caption{The key rate (in bits per pulse) resulting from the MDI protocol  using the homodyne detections for an asymmetric setting consistent with the satellite directly overhead Alice (with Alice-Bob and Alice-satellite distances set at 1000km). Also shown is the impact of the reconciliation efficiency for the same asymmetric settings ($v=60$, and ${\varepsilon _A} = {\varepsilon _B} = 0.02$).}\label{fig:2}
    \end{center}
\end{figure}

 It is perhaps interesting to compare our CV-MDI protocol with a scheme in which the satellite acts as a reflecting relay (trustworthy relay). This scheme referred to as the \emph{direct transmission scheme} has been studied in our previous work \cite{Neda1} in terms of Gaussian entanglement generation. In this scheme, Alice initially possesses a TMSV state, where one mode is held by Alice, and the other mode is transmitted towards Bob via a reflecting satellite over two independent fading channels. After the distribution of entanglement, Alice and Bob can proceed with CV-QKD by applying homodyne (or heterodyne) detections to their own modes. In the typical point-to-point CV-QKD, the reverse reconciliation (RR) scenario always leads to higher key rates than the direct reconciliation (DR) scenario \cite{RR_2002}. Hence, we only consider the RR scenario where Bob's data is the reference of reconciliation. We again consider the symmetric setting, i.e., ${\sigma _{b\_A}} = {\sigma _{b\_B}}=\sigma _{b}$, and the same type of detection by the trusted parties.

 Fig.~1(a) also shows the key rate  of the direct transmission scheme for two types of detection, homodyne detection and heterodyne detection, as a function of the total mean fading losses. Note that for these calculations the value of the quadrature variance is $v=60$, and the value of the excess noise (the receiver noise at Bob's station) is chosen the same as that in the CV-MDI protocol.
These results, as expected, illustrate that the MDI protocol is not as effective as the direct transmission scheme in terms of the quantum key rates.  The difference in the key rates shown between the direct transmission protocol and the MDI protocol simply illustrates the cost in moving from a trustworthy relay to an untrustworthy relay.

We have also simulated the performance of the MDI protocol in the more generic asymmetric settings anticipated for satellite communications
(i.e. one link has less fading loss than the other).  For such circumstances, in general we find similar trends to those shown in Fig.~1(a),
 but with slightly higher key rates when the trusted party having the lower loss is chosen as the reference of reconciliation. An example of such a calculation is shown in Fig.~2 for ${\sigma _{b\_A}} = 0.54{\sigma _{b\_B}}$. Such a setting simulates the circumstance when the satellite is directly overhead one of the ground stations. From Fig.~2 we can see that the impact of the geometry is very small, with the key rate improvement only at  about $20\%$ at low losses and almost negligible at the higher losses of 25-30dB. Such a conclusion on the relevance of the geometry is not significantly changed unless the ratio ${\sigma _{b\_A}}/{\sigma _{b\_B}}$  is set at very unrealistic values.\footnote{That is, unrealistic in the context of a satellite being the measurement device between two ground stations. Of course low ratios of ${\sigma _{b\_A}}/{\sigma _{b\_B}}$ can be anticipated if one of the ``ground" stations is replaced with a high-flying aircraft or a space-based transmitter. In such cases, order of magnitude improvements in the quantum key rates can be anticipated.}
In Fig.~2 we also take the opportunity to study the influence of  the reconciliation efficiency $\xi$ alluded to earlier. Thus far we have assumed a perfect reconciliation, $\xi=1$. The lower curves of Fig.~2 show the anticipated key rates for a realistic high (but achievable) value of $\xi=0.95$, and for comparison an inefficient value $\xi=0.8$. These results indicate the importance of efficient classical codes (such as well-designed LDPC codes) in the reconciliation phase.

The results shown in Figs.~1-2 can be translated into rates per second after multiplication by the anticipated pulse generation rate. Pulse rates of order 100MHz can be anticipated by state-of-the-art systems. The average key rate, of course, must be determined through a weighted  integration over all possible geometrical configurations. However, from the results presented in Figs.~1-2
we already see that, dependent on the scheme adopted and on the actual loss rates, quantum key rates in the range of up to a few kbit/s can be anticipated for space-based implementations of the MDI protocol.

 At the cost of additional complexity at the satellite, other variants of the MDI paradigm are possible. For example, inclusions of non-Gaussian operations at the relay, such as a single-photon subtraction from each incoming pulse, prior to the Bell measurement could be invoked. Such photonic subtraction would generate a conditional state between Alice and Bob that would be  non-Gaussian in nature, possibly possessing enhanced entanglement - and therefore leading to higher key rates. Our preliminary analysis of such a single-photon subtraction scheme appears to indicate that higher key rates  will not be forthcoming - largely because of the low probabilities associated with the photon subtraction procedure. Nonetheless, future work in this area could explore this issue in more detail, possibly extending the scope to include even more exotic quantum operations at the relay.  It is certainly not impossible that the quantum key generation rates presented here can be significantly improved upon if more sophisticated quantum operations are possible on board the satellite.

\section{Conclusion}
In this work we have explored a CV-based MDI protocol for two terrestrial stations communicating via an untrustworthy LEO satellite. In terms of the quantum key rates, we have found that the probabilistic nature of the fading channels towards the satellite can lead to significant and practical advantages relative to fixed-attenuation channels. Specifically, we found that the MDI protocol is feasible over high-loss fading channels, while it is only able to generate non-zero quantum key rates over low-loss fixed-attenuation channels.

The results given here represent a valuable quantitative assessment of measurement-device-independence technology as applied to future space-based quantum communications. They are particularly interesting given the outcome that ultra-secure communications between two ground stations can be made viable even when the relay satellite is held by an adversary. The best an adversary can do in such circumstances is to report false measurement outcomes - an act which provides her with no insight to the actual key. Such false reporting is detectable and can only reduce the final key rate generated by the two ground stations.

\appendix

An equivalence between the PM schemes and the EB schemes has  been used in this analysis. The path to showing such equivalence has been previously detailed in \cite{CV_MDI2} for the case where the initial quantum resource is a coherent state. Here we detail the similar path to equivalence when the initial quantum state is a squeezed state.

 To make progress we define a \emph{modified PM scheme} in which  Alice (Bob), initially possesses the TMSV state $\rho_{12}$ ($\rho_{34}$), and then makes a homodyne detection on mode~1 (mode~4). It is straightforward to show that such a modified PM scheme is equivalent to the PM scheme with a squeezed state as the initial quantum resource, and we simply adopt such an equivalence here.
% As such, equivalence between the modified PM scheme and the EB scheme also demonstrates equivalence between the PM scheme and the EB scheme.
 To show equivalence between the modified PM scheme and the EB scheme,
 we demonstrate that the probabilities for generic data outcomes for both schemes are identical.

 In the modified PM scheme, the initial density matrix of the system including Eve, $\rho_E$, can be written as ${\rho _0} = {\rho _{12}} \otimes {\rho _E} \otimes {\rho _{34}}$. We assume Alice and Bob measure the $\hat q$ quadrature of modes~1 and 4, yielding $q_1$ and $q_4$ with the probability $p(q_1,q_4)$. After their measurement, the density operator of the system is given by
\begin{equation}\label{1}
{\rho _{23E}} = \frac{{\left\langle {{q_1},{q_4}} \right|{\rho _0}\left| {{q_1},{q_4}} \right\rangle }}{{p\left( {{q_1},{q_4}} \right)}}.
\end{equation}
Modes~2 and 3 are then transmitted towards the untrusted relay, in which the Bell measurement is applied to the incoming modes, yielding $q_s$ and $p_t$ with the probability $p(q_s,p_t)$. Here, we assume any transformation on the transmitted modes~2 and 3 as well as Eve's ancillas (before obtaining the outcomes $q_s$ and $p_t$) can be inserted into a global unitary operator $U_{23E}$. Thus, the probability $p_m$ of all the measurement outcomes $\left( {{q_1},{q_4},{q_s},{p_t}} \right)$ is given by % defined by $p_j$
\begin{equation}\label{2}
\begin{array}{l}
p_m\left( {{q_1},{q_4},{q_s},{p_t}} \right) =\\

\,\,\,\,\,\,\,\,\,\,\,\,\,\,\,\,\,\,\,\,\,\,\,\,\,\,\,\,\,\,\,\,p\left( {{q_1},{q_4}} \right)\left\langle {{q_s},{p_t}} \right|{U_{23E}}\,\,{\rho _{23E}}\,\,U_{23E}^\dag \left| {{q_s},{p_t}} \right\rangle.
\end{array}
\end{equation}
Substituting $\rho _{23E}$ of Eq.~\eqref{1} into Eq.~\eqref{2}, we will have
\begin{equation}\label{3}
\begin{array}{l}
p_m\left( {{q_1},{q_4},{q_s},{p_t}} \right) =\\

\,\,\,\,\,\,\,\,\,\,\,\,\,\,\,\,\,\,\,\,\,\,\,\,\,\,\,\,\,\,\,\,\left\langle {{q_1},{q_4},{q_s},{p_t}} \right|{U_{23E}}\,\,{\rho _0}\,\,U_{23E}^\dag \left| {{q_1},{q_4},{q_s},{p_t}} \right\rangle .
\end{array}
\end{equation}
The Bell measurement result $(q_s,p_t)$ is then communicated to Alice and Bob. According to the result $(q_s,p_t)$, Bob modifies his data $q_4$ by $Q_4=q_4+kq_s$, where $k$ is the gain factor in the modified PM scheme. However, Alice keeps her data unchanged, i.e., $Q_1=q_1$. Thus, the probability $p$ of the final data $\left( {{Q_1},{Q_4},{q_s},{p_t}} \right)$ is given by
\begin{equation}\label{4}
\begin{array}{l}
 p\left( {{Q_1},{Q_4},{q_s},{p_t}} \right) = p_m\left( {{Q_1},{Q_4} - k{q_s},{q_s},{p_t}} \right) = \\
\\
\left\langle {{Q_1},{Q_4} - k{q_s},{q_s},{p_t}} \right|{U_{23E}}{\rho _0}U_{23E}^\dag \left| {{Q_1},{Q_4} - k{q_s},{q_s},{p_t}} \right\rangle  = \\
\\
\left\langle {{Q_1},{Q_4},{q_s},{p_t}} \right|D\left( {k{q_s}} \right){U_{23E}}{\rho _0}U_{23E}^\dag {D^\dag }\left( {k{q_s}} \right)\left| {{Q_1},{Q_4},{q_s},{p_t}} \right\rangle ,
\end{array}
\end{equation}
where $D\left( {k{q_s}} \right)$ is the displacement operator. Note, if Bob measures the $\hat p$ quadrature of his mode~4, yielding $p_4$, after receiving the Bell measurement result, he modifies his data as $P_4 =p_4-kp_t$.

Now, we consider the  EB scheme, where Alice and Bob generate the TMSV states $\rho_{12}$ and $\rho_{34}$ respectively, and transmit modes~2 and 3 towards the untrusted relay. After the Bell measurement at the relay with the outcome $(q_s,p_t)$ and the probability $p(q_s,p_t)$, the density operator of the system is given by
\begin{equation}\label{5}
{\rho _{14}} = \frac{{\left\langle {{q_s},{p_t}} \right|{U_{23E}}\,\,{\rho _0}\,\,U_{23E}^\dag \left| {{q_s},{p_t}} \right\rangle }}{{p\left( {{q_s},{p_t}} \right)}},
\end{equation}
where the initial density matrix of the system, $\rho_0$, and the global unitary operator $U_{23E}$ are the same as those in the modified PM scheme. After broadcasting the Bell measurement results $(q_s,p_t)$ to Alice and Bob, Alice keeps mode~1 unchanged, while Bob displaces mode~4 by the displacement operator ${D}\left( {g{q_s}} \right)$, where $g$ is the gain factor in the EB scheme. At this step the density operator of the system is given by
\begin{equation}\label{6}
{\rho _{14'}} = \frac{{\left\langle {{q_s},{p_t}} \right|D\left( {g{q_s}} \right){U_{23E}}\,\,{\rho _0}\,\,U_{23E}^\dag {D^\dag }\left( {g{q_s}} \right)\left| {{q_s},{p_t}} \right\rangle }}{{p\left( {{q_s},{p_t}} \right)}},
\end{equation}
where mode~$4'$ is mode~4 after the displacement. Then Alice and Bob apply the homodyne detections on mode~1 and $4'$, respectively. Here, again we assume Alice and Bob measure the $\hat q$ quadrature of modes~1 and $4'$, yielding $Q_1$ and $Q_4$. The probability $p$ of the data $(Q_1,Q_4)$ given the Bell measurement result $(q_s,p_t)$ is
\begin{equation}\label{6}
p\left( {{Q_1},{Q_4}\left| {{q_s},{p_t}} \right.} \right) = \left\langle {{Q_1},{Q_4}} \right|{\rho _{14'}}\left| {{Q_1},{Q_4}} \right\rangle.
\end{equation}
Thus, the probability of the final data $\left( {{Q_1},{Q_4},{q_s},{p_t}} \right)$ is given by
\begin{equation}\label{7}
\begin{array}{l}
 p\left( {{Q_1},{Q_4},{q_s},{p_t}} \right) = p\left( {{Q_1},{Q_4}\left| {{q_s},{p_t}} \right.} \right)p\left( {{q_s},{p_t}} \right) = \\
\\
\left\langle {{Q_1},{Q_4},{q_s},{p_t}} \right|D\left( {g{q_s}} \right){U_{23E}}{\rho _0}U_{23E}^\dag {D^\dag }\left( {g{q_s}} \right)\left| {{Q_1},{Q_4},{q_s},{p_t}} \right\rangle .
\end{array}
\end{equation}
Comparing Eqs.~\eqref{4} and \eqref{7}, the probability of the final data is the same for the modified PM scheme and the EB scheme when $k=g$. As such, the modified PM scheme and our EB scheme are equivalent. The equivalence between the PM scheme and the modified PM scheme leads to the equivalence claimed in this paper.

\end{document}